# Stability Analysis of LQR-ANFIS Control Schemes on 2-degree-of-freedom Inverted Pendulum Systems.

Shamanth S*, Aditya Kumar Chari, and Harshitha S

**Abstract:** The concepts of stability and balance represent many critical problems faced by engineering today. The inverted pendulum on a cart is one such non-linear, unstable, multivariate system whose goal is to determine a suitable control action given to the cart such that it stabilizes the pendulum in an upright vertical position. This paper therefore, aims to design and study a highly robust MISO control structure using Linear Quadratic Regulation, Fuzzy logic and Neural Networks called Two-Stage LQR-based-ANFIS (referred to as TS-LA) for the stabilization of Inverted Pendulums. The proposed controller is implemented on a Simulink model of the Inverted Pendulum constructed through relevant mathematical and state space modelling using Newtonian and Lagrangian mechanics. Applying external disturbances, transient parameters are obtained and are benchmarked against standard conventional controllers to perform comparative analysis and showcase its disturbance rejection capabilities.

## 1. INTRODUCTION

The inverted pendulum on a cart is an important problem in control theory and is used frequently for benchmarking control system performance. Being a highly unstable, non-linear, time variant system, it consists of a movable cart with one of the pendulum's ends pivoted on it, with the other being free. Control signal or force is given to the cart through an actuator such as a servo motor, and hence due to moment of inertia the angle of the pendulum is also affect ted. This way, by precise movement of the cart, one can stabilize the pendulum in an upright position. Therefore such an unstable process necessitates the need for control systems that always ensure good stability and tracking performance of the plant's state regardless of any uncertain parametric variation.

In that regard, conventional controllers exhibit excellent performance given that the mathematical model of the plant is precisely defined. This means that there are inevitable performance trade-offs with uncertainty in the model. On the other hand, adaptive controllers adjust their control action to the present state of the plant, improving themselves in case of any adverse, unforeseen events making them robust under non-linearity. There have been numerous demonstrations of this since the past few decades, highlighting the technological shift from conventional to more intelligent adaptive techniques.

An example of one such intelligent approach is the implementation of fuzzy theory, first formulated and published by Dr. Zadeh [1]. In literature there have been numerous instances where motors and actuators have been interfaced and controlled with fuzzy logic controllers, to overcome the performance degradation faced by conventional controllers due to non-linearities [2-5]. An early example of this is the work done by Liaw and Wang (1991) where a high performance induction motor drive was controlled based on limit-cycle control technique. The controller dynamics were constructed using human experience of the system [6]. Subsequent later developments in the field included hybrid fuzzy-PID structures that bridged the gap between adaptive and conventional controllers by allowing the PID controller to self-tune its gains using fuzzy sets [7-11]. The problem of stabilizing an inverted pendulum was also addressed by a non-linear hybrid controller as S. Malik and S. Mishra (2015) proposed the use of such a fuzzy-PID controller for the stabilization of a pendulum on a cart and observed its superiority over fixed-gain PID [12]. The drawback observed with this method however, is rule explosion in multivariable systems. The extensive number of rules required which are mostly manually tuned though trial and error even if expert knowledge is involved (rule explosion). On the other hand, Linear Quadratic

Shamanth S*, Aditya Kumar Chari and Harshitha S are with JSS Academy of Technical Education, JSSATE-B Campus, Dr. Vishnuvardhan Road, Srinivaspura, Bengaluru, Karnataka 560060, India. (e-mails: shamanthsreekanth@gmail.com, adityachari67@gmail.com, harshithashivaprasad@gmail.com).
\* Corresponding author.

Regulators (LQR) have been designed using the more convenient state space modelling to address the issue of achieving an optimal target solution, through control laws [13-15], following the design of neuro-fuzzy techniques in the 1990s which saw the introduction of Adaptive Neuro Fuzzy Inference Systems [16-18]. The centralized nature and the parallel processing capability of ANFIS to generate FIS (Fuzzy Inference Systems) on its own provided a massive advantage over previously designed intelligent control techniques [19]. Even in recent times, several works have demonstrated the superiority of fuzzy based hybrid controllers with Parallel kinematic machine (PKMs) control [20], small satellite attitude control [21], etc. Meta-heuristic techniques like Particle Swarm Optimization (PSO) and expert systems like Model Predictive Control (MPC) have also gained traction with motor control [22-23], and the standard PI control has been proven too simple for complex non-linear applications [24]. Considering these examples in literature, this paper attempts to further design and simulate "TS-LA", an LQR based ANFIS system over two stages, for the stabilization and balance of a 2-degree of freedom inverted pendulum on a cart. The physical structure of the system is modelled through a subsystem implemented on Simulink. LQR controllers are first designed using state space modelling. Subsequent data is then logged to create a database. An ANFIS network is then formulated in MATLAB which is trained based on the data obtained.

The paper is organized as follows: Section 2 breaks down the mathematical modelling of the system. Section 3 highlights the design of the LQR based training model and how data is formulated. Section 4 details the design of the ANFIS controller and their subsequent training procedures. Section 5 deals with the simulation results and finally section 6 draws conclusions with scope for future work.

## 2. MATHEMATICAL MODELLING

Mathematical modelling that deals with the kinematics of an object is crucial as the equations that follow govern the motion of a physical entity. This includes variables like position, velocity and acceleration.

### 2.1. Equations of motion

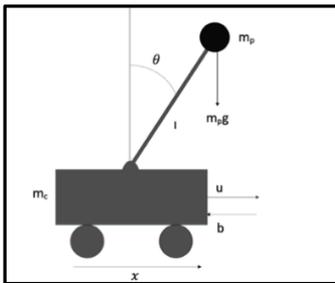

Fig. 1. Free body diagram of an inverted pendulum

The different displacements taking place in this system are: i) horizontal displacement of the cart, ii) horizontal displacement of the pendulum, iii) vertical displacement of the pendulum and iv) angular displacement of the pendulum.

The horizontal displacement of the cart is given by control input and the horizontal component of the cart,

$$m_c \frac{d^2x}{dt^2} = u - H - b\frac{dx}{dt} \quad (1)$$

The horizontal displacement of the pendulum is given by the horizontal distance travelled by its centre of mass and the displacement of the cart,

$$m_p \frac{d^2}{dt^2}(x + l\sin\theta) = H \quad (2)$$

The vertical displacement of the pendulum is given by vertical distance travelled by its centre of mass, weight of the pendulum in the downwards direction and the vertical component exerted by the cart on the pendulum,

$$m_p \frac{d^2}{dt^2}(l - l\cos\theta) = m_p g - V \quad (3)$$

i.e,

$$m_p \frac{d^2}{dt^2}(-l\cos\theta) = m_p g - V \quad (4)$$

as the derivative of $l$ is 0.

Finally, the rotational displacement of the pendulum is given by the horizontal and vertical displacements and the moment of inertia $J$.

$$J \frac{d^2\theta}{dt^2} = V\sin\theta l - H\cos\theta l \quad (6)$$

A dynamical system given by f(x) has equilibrium points at x=0 where the derivative is zero, $f(x_0) = 0$. Based on how the system performs around x 0, they are stable or unstable. Inverted pendulum systems have stable and unstable equilibria.

By linearising the system around its unstable upright equilibrium, a simpler linear model of the system is creating along with its transfer functions around this point.

Taking $\theta = \pi$, to eliminate the non-linearities $\sin\theta$ and $\cos\theta$, we can assume that for small values of $\theta$,

$$\cos\theta \approx -1 \quad (6)$$
$$\sin\theta \approx -\theta \quad (7)$$
$$\frac{d^2\theta}{dt^2} \approx 0 \quad (8)$$

i.e., equations (6)-(8) become

$$m_p \frac{d^2}{dt^2}(x - l\theta) = H \quad (9)$$
$$0 = m_p g - V \quad (10)$$
$$J \frac{d^2\theta}{dt^2} = -V\theta l + Hl \quad (11)$$

Substituting $V$ and $H$ from (9)-(10) in (11), we have the relation

$$J \frac{d^2\theta}{dt^2} = -m_p g\theta l + m_p \frac{d^2}{dt^2}(x - l\theta)l \quad (12)$$
$$(J + m_p l^2)\frac{d^2\theta}{dt^2} - m_p l\frac{d^2x}{dt^2} + m_p g l\theta = 0 \quad (13)$$

Substituting $H$ from (9) in (1), we get

$$m_c \frac{d^2x}{dt^2} = u - m_p \frac{d^2}{dt^2}(x - l\theta) - b\frac{dx}{dt} \quad (14)$$
$$(m_c + m_p)\frac{d^2x}{dt^2} + b\frac{dx}{dt} - m_p l\frac{d^2\theta}{dt^2} = u \quad (15)$$

Therefore, equations (13) and (15) represent the equations of motion for the linearized model of the inverted pendulum system at equilibrium. They can also be written as:

$$(J + m_p l^2)\ddot{\theta} - m_p l\ddot{x} + m_p g l\theta = 0 \quad (16)$$
$$(m_c + m_p)\ddot{x} + b\dot{x} - m_p l\ddot{\theta} = u \quad (17)$$

### 2.1.2. Lagrangian approach:

Cross verifying the equations (16) and (17) is possible using another way to understand the equations of motion without explicitly using Newton's laws, through Lagrangian mechanics. It uses the principle of least action, which makes for a more axiomatic theory. The general form of the Lagrange equation is given as:
$$L = K - P \tag{18}$$
Where $L$, the Lagrangian is defined as the difference between the kinetic and potential energy of a system. It is important to note that we use classical physics in this context and not relativistic mechanics. It is also important to note that the Lagrangian is mathematical quantity and not a physical quantity unlike the Hamiltonian which represents the total energy of the system. In this case, the kinetic energy of the system, $K$, is given by combined kinetic energies of the cart and the pendulum, i.e.,
$$K = \tfrac{1}{2}m_c v^2 + \tfrac{1}{2}m_p v^2 + \tfrac{1}{2}J\omega^2 \tag{19}$$
Where, $\omega = \dot{\theta}$ which is the first derivative of the angle of the pendulum giving angular velocity and $v = \dot{x}$ which is the first derivative of the position of the cart giving cart velocity.

$K$ can be expanded to get:
$$K = \tfrac{1}{2}m_c \dot{x}^2 + \tfrac{1}{2}m_p \left(\tfrac{d}{dt}(x + l\sin(\theta))\right)^2 +$$
$$\tfrac{1}{2}m_p \left(\tfrac{d}{dt}(-l\cos(\theta))\right)^2 + \tfrac{1}{2}J\dot{\theta}^2 \tag{20}$$
$$K = \tfrac{1}{2}(m_c + m_p)\dot{x}^2 + \tfrac{1}{2}m_p l^2 \dot{\theta}^2 + m_p l\cos(\theta)\dot{x}\dot{\theta} + \tfrac{1}{2}J\dot{\theta}^2 \tag{21}$$

The potential energy of the system, P, is given by:
$$P = m_p g l\cos(\theta) \tag{22}$$
The total Lagrange L, therefore becomes,
$$L = \tfrac{1}{2}(m_c + m_p)\dot{x}^2 + \tfrac{1}{2}(J + m_p l^2)\dot{\theta}^2 + ml\cos(\theta)\dot{x}\dot{\theta} - m_p g l\cos(\theta) \tag{23}$$

### 2.1.3. Euler-Lagrange:

Following the calculus of variations, the Euler-Lagrange equations, shown in equations (24) and (25), are fundamental equations for solving optimization problems like the stability of an inverted pendulum system where a differentiable function has stationary points.
$$\frac{d}{dt}\left(\frac{\partial L}{\partial \dot{\theta}}\right) - \frac{\partial L}{\partial \theta} = 0 \tag{24}$$
$$\frac{d}{dt}\left(\frac{\partial L}{\partial \dot{x}}\right) - \frac{\partial L}{\partial x} = u \tag{25}$$

As the motion of the inverted pendulum system hinges on the position of the cart $x$ and the angle of the pendulum $\theta$, we can assume them to be the generalized coordinates that determine the configuration of the inverted pendulum system and how the coordinates change as functions of time.
$$\frac{\partial L}{\partial x} = 0 \tag{26}$$
$$\frac{\partial L}{\partial \dot{x}} = (m_c + m_p)\dot{x} + m_p l\cos(\theta)\dot{\theta} \tag{27}$$
$$\frac{\partial L}{\partial \theta} = -m_p l\sin(\theta)\dot{x}\dot{\theta} + m_p g\sin(\theta) \tag{28}$$
$$\frac{\partial L}{\partial \dot{\theta}} = (J + m_p l^2)\dot{\theta} + m_p l\cos(\theta)\dot{x} \tag{29}$$

Substituting equations (26)-(29) in equations (24) and (25),
$$\frac{d}{dt}\left((J + m_p l^2)\dot{\theta} + m_p l\cos(\theta)\dot{x}\right) - \left(-m_p l\sin(\theta)\dot{x}\dot{\theta} + m_p g\sin(\theta)\right) = 0 \tag{30}$$
$$\frac{d}{dt}\left((m_c + m_p)\dot{x} + m_p l\cos(\theta)\dot{\theta}\right) - 0 = u \tag{31}$$

Solving the ODEs, we obtain,
$$(J + m_p l^2)\ddot{\theta} + m_p l\cos(\theta)\ddot{x} - m_p g l\sin(\theta) = 0 \tag{32}$$
$$(m_c + m_p)\ddot{x} + b\dot{x} + m_p l\cos(\theta)\ddot{\theta} - m_p \sin(\theta)\dot{\theta}^2 = u \tag{33}$$

Linearizing equations (32) and (33) around $\theta = \pi$, we get equations (34) and (35), which are the same as equations (16) and (17), proving their validity.
$$(J + m_p l^2)\ddot{\theta} - m_p l\ddot{x} + m_p g l\theta = 0 \tag{34}$$
$$(m_c + m_p)\ddot{x} + b\dot{x} - m_p l\ddot{\theta} = u \tag{35}$$

### 2.2. Transfer Function:

To obtain transfer functions of a system, we need to run the equations of motions that govern the said system through Laplace transforms that convert simple variables from time domain to complex variables in the s-plane or frequency domain. The general formula for Laplace transform is given by:
$$F(s) = \int_0^\infty f(t)e^{-st}\,dt \tag{36}$$
By using this on equations (13) and (15), we get
$$(m_p l^2 + J)s^2\theta(s) + m_p l s^2 X(s) + m_p g l\theta(s) = 0 \tag{37}$$
$$(m_c + m_p)s^2 X(s) + m_p l s^2 \theta(s) = U(s) - bsX(s) \tag{38}$$

A transfer function only represents the relationship between a single input and a single output at a time. Therefore, we divide the system into two subcomponents, the pendulum and the cart respectively and first find the transfer function of the pendulum. To find the first transfer function for the output $\theta(s)$ and an input of $U(s)$, we need to eliminate $X(s)$ from the above equations. We get,
$$\frac{\theta(s)}{U(s)} = \frac{\frac{m_p l}{\alpha}s}{s^3 + \frac{b(J + m_p l^2)}{\alpha}s^3 - \frac{(m_c + m_p)m_p g l}{\alpha}s^2 - \frac{bm_p g l}{\alpha}s} \tag{39}$$

Likewise, to find the second transfer function for the output $X(s)$ and an input of $U(s)$, we need to eliminate $\theta(s)$ from equations (37) and (38). We get,
$$\frac{\theta(s)}{U(s)} = \frac{\frac{m_p l}{\alpha}s}{s^3 + \frac{b(J + m_p l^2)}{\alpha}s^3 - \frac{(m_c + m_p)m_p g l}{\alpha}s^2 - \frac{bm_p g l}{\alpha}s} \tag{40}$$
Where $\alpha$ is $\left[(m_c + m_p)(J + m_p l^2) - (m_p l)^2\right]$

### 2.3. State space analysis

The complete system model for a linear continuous time system consists of (i) a set of $n$ state equations, defined in terms of the matrices A and B, and (ii) a set of output equations that relate any output variables of interest to the state variables and inputs, and expressed in terms of the C and D matrices. The task of modelling the system is to derive the elements of the matrices, and to write the system model in the form:
$$\dot{x} = Ax + Bu \tag{41}$$
$$y = Cx + Du \tag{42}$$
Taking,
$$\begin{pmatrix} x \\ \dot{x} \\ \theta \\ \dot{\theta} \end{pmatrix} = \begin{pmatrix} x_1 \\ x_2 \\ x_3 \\ x_4 \end{pmatrix} \rightarrow \begin{pmatrix} \dot{x} \\ \ddot{x} \\ \dot{\theta} \\ \ddot{\theta} \end{pmatrix} = \begin{pmatrix} \dot{x}_1 \\ \dot{x}_2 \\ \dot{x}_3 \\ \dot{x}_4 \end{pmatrix} \tag{43}$$
the state space representation for the inverted pendulum system can be expressed in the form of equations (44) and (45):

$$\begin{pmatrix} \dot{x}_1 \\ \dot{x}_2 \\ \dot{x}_3 \\ \dot{x}_4 \end{pmatrix} = \begin{pmatrix} 0 & 1 & 0 & 0 \\ 0 & a_1 & a_2 & 0 \\ 0 & 0 & 0 & 1 \\ 0 & a_3 & a_4 & 0 \end{pmatrix} \begin{pmatrix} x \\ \dot{x} \\ \theta \\ \dot{\theta} \end{pmatrix} + \begin{pmatrix} 0 \\ a5 \\ 0 \\ a6 \end{pmatrix} u \quad (44)$$

$$y = \begin{pmatrix} 1 & 0 & 0 & 0 \\ 0 & 0 & 1 & 0 \end{pmatrix} \begin{pmatrix} x \\ \dot{x} \\ \theta \\ \dot{\theta} \end{pmatrix} + \begin{pmatrix} 0 \\ 0 \end{pmatrix} u \quad (45)$$

$a_1, a_2, a_3, a_4, a_5, a_6$ are arbitrary variables taken for simplicity, where:

$$a_1 = \frac{-(m_p l^2 + J)b}{(m_c + m_p)(m_c l + J) - (m_p l)^2} \quad (46)$$

$$a_2 = \frac{m_p{}^2 g l^2}{(m_c + m_p)(m_c l + J) - (m_p l)^2} \quad (47)$$

$$a_3 = \frac{m_p l b}{(m_c + m_p)(m_c l + J) - (m_p l)^2} \quad (48)$$

$$a_4 = \frac{-(m_c + m_p) m_p g l}{(m_c + m_p)(m_c l + J) - (m_p l)^2} \quad (49)$$

$$a_5 = \frac{J + m_p l^2}{(m_c + m_p)(m_c l + J) - (m_p l)^2} \quad (50)$$

$$a_6 = \frac{-m_p l}{(m_c + m_p)(m_c l + J) - (m_p l)^2} \quad (51)$$

### 2.4. Root Locus

For the described inverted pendulum system, there is a need to know if the system is BIBO stable and the degree to which this system must be adjusted or compensated to produce the desired performance characteristics. Hence, two root locus plots are obtained for the transfer functions defined by equations 39 and 40 in MATLAB, which are illustrated in Fig. 2.

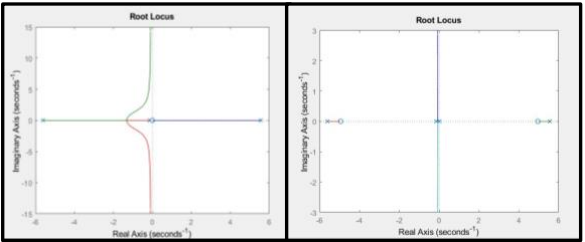

Fig. 2. Root locus analysis of the transfer function of the cart and the pendulum.

In Fig. 2. the root locus plot of the third order system shows that we have open loop complex conjugate poles s = 5.5651, -5.6041, -0.1428. The root locus plot of the fourth order system shows that we have open loop complex conjugate poles at s = 0, s = 5.5651, s = -5.6041, s = -0.1428. As we can see, there is a real, positive pole on the right-hand side for both graphs. As the right-hand side of the s-domain is the unstable region, the system is unstable as well.

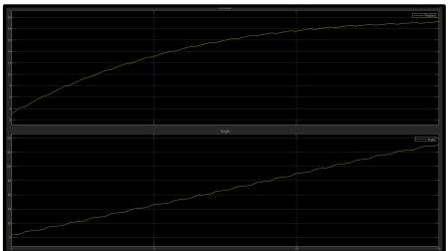

Fig. 3. Open loop response analysis to a small-scale impulse disturbance.

The open loop system does not follow BIBO stability criteria, as any small bounded input given will produce an unbounded output whose energy magnitude to infinity. This is observed in Fig. 3. where a small impulse disturbance is given the open loop inverted pendulum system modelled as shown in Fig. 7. making the cart position and angle of the pendulum grow unbounded. This proves there is a need for the unstable system to be controlled with a closed loop system.

### 2.5. Controllability

In order to be able to perform the requirements with the given dynamic system under control input, the system must be controllable. In this linear, continuous time system represented by equation (41), controllability is tested using a controllability matrix $Co$ defined by:

$$Co(A, B) = [B : AB : A^2B : \dots : A^{n-1}B] \quad (52)$$

Where A is the system matrix of size and B is the input matrix. The system is said to be controllable if the rank of $Co$ is the same as the rank of the state transition matrix $A$, or if the determinant of the controllability matrix is non-zero. For the given dynamic system under consideration, we see that

$$Co(A,B) = \begin{pmatrix} 0 & 1.8182 & -0.3306 & 12.2089 \\ 1.8182 & -0.3306 & 12.2089 & -4.4287 \\ 0 & 4.5455 & -0.8264 & 141.8858 \\ 4.5455 & -0.8264 & 141.8858 & -31.3196 \end{pmatrix} \quad (53)$$

The rank of $Co(A, B)$ is 4, and the determinant is non-zero and positive. Hence the system is observed to be controllable.

### 3. CONTROL DESIGN

#### 3.1. Overview:

The control operation is carried out over two stages as shown in Fig. 4 and 5. The first stage uses LQR to take four inputs, cart position $x$, cart velocity $\dot{x}$, angle of the pendulum $\theta$, and angular velocity of the pendulum $\dot{\theta}$. The LQR algorithm performs control action such that the error is driven to global minimum based off the equilibrium position of $x = 0$ and $\theta = \pi$. The resulting data including the control output $u$ is data logged into a database that is carried over to stage 2, where it is fed to an ANFIS controller. A fuzzy inference system is automatically generated and trained using Artificial Neural Networks (ANNs), and the trained ANFIS model can be implemented to perform actual control action on the subsystem.

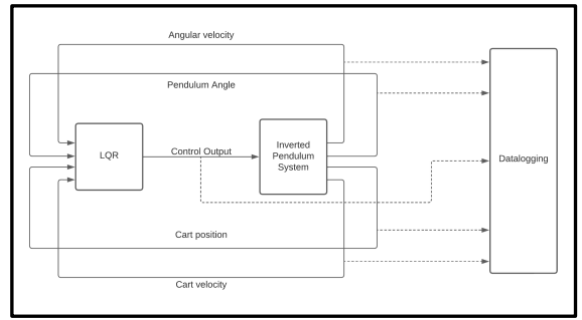

Fig. 4. Stage 1 using LQR

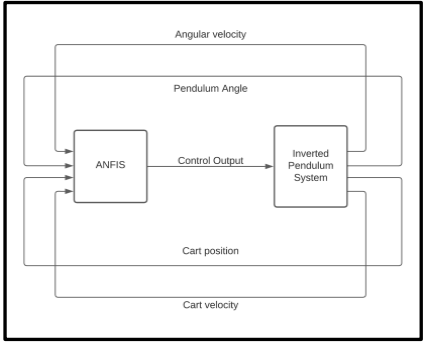

Fig. 5. Stage 2 using ANFIS

### 3.2. LQR

The dynamic system under consideration is modelled in state space representation as given in equations 22 and 23. The behaviour of the system is characterised and influenced by the eigenvalues of control matrix $A$, which are the poles of the system. The control input $u(t)$ therefore needs to work on the actuator such that these poles are shifted. LQR or Linear Quadratic Regulator is a control technique that works exclusively on these state space matrices, altering the states of the system. Therefore it is a type of full state feedback controller. Consider a system in state-space form given by equation (41). The control action in LQR is

$$u(t) = -Kx(t) \tag{54}$$

Where $x(t)$ is the state vector and $K$ is the gain given to it. LQR is based on the optimal control theory where, as per definition, system is operated with optimized parameters as per requirement. These optimal parameters are the solution set of an infinite horizon quadratic cost function called the LQ problem.

$$CF = \int_0^\infty [x^T Q x + u^T R u]\, dt \tag{55}$$

Here, $Q$ and $R$ are matrices that determine the weights on the states and the weights on the control input given respectively, where $Q = Q^T \geqslant 0$ and $R = R^T \geqslant 0$.

The goal is to find the optimal cost-to-go or value function $J^*(x)$ which satisfies the Hamilton-Jacobi-Bellman equation

$$0 = \min\left[x^T Q x + u^T R u + \frac{\partial J^*}{\partial x}(Ax + Bu)\right] \tag{56}$$

We can verify that the cost-to-go function is quadratic by choosing the form:

$$J^*(x) = x^T S x \tag{57}$$

Where $S = S^T \geqslant 0$. The gradient of this function is

$$\frac{\partial J^*}{\partial x} = 2x^T S \tag{58}$$

Therefore, we observe that the terms in equation 30 are quadratic and convex in nature. Taking the solution where the gradient of these terms vanishes,

$$\frac{\partial}{\partial u} = 2u^T R + 2x^T SB = 0 \tag{59}$$

This further yields the optimal policy,

$$u^* = \pi^*(x) = -R^{-1}B^T S x = -Kx \tag{60}$$

Insertinxg this in the Hamilton-Jacobi-Bellman,

$$0 = x^T[Q - SBR^{-1}B^T S + 2SA]x \tag{61}$$

$2SA$ is not symmetric in nature but since $x^T SAx = x^T A^T S x$, we can write

$$0 = x^T[Q - SBR^{-1}B^T S + SA + A^T S]x \tag{62}$$

Since this condition must hold true for all values of $x$, we can consider the matrix equation

$$0 = SA + A^T S - SBR^{-1}B^T S + Q \tag{63}$$

Equation (63) is a version of the algebraic Riccati equation, upon solving which yields the $K$ term in equation (64) used in equation (54).

$$K = -R^{-1}B^T S \tag{64}$$

### 3.3. ANFIS

Assume the ANFIS under consideration has $p$ inputs $(Z_{t-1}, Z_{t-2} \ldots Z_{t-p})$ and one output $Z_t$, with $m$ rules. If $Z_{t-k}$ is $A_{kj}$, then

$$Z_t^{(j)} = \theta_{j0} + \sum_{k=1}^p \theta_{jk} Z_{t-k} \tag{65}$$

With, $\theta_{j0}$ and $\theta_{jk}$ as linear parameters, $A_{kj}$ as non-linear parameters where, $j = 1,2,3 \ldots m$ and $k = 1,2,3 \ldots p$. It typically consists of 5 layers as shown in Fig. 7.

The first layer consists of adaptive nodes and is responsible for allocating membership functions to discrete input sets for fuzzification.

The second layer generates fuzzy rules and assigns firing strength to each rule using AND operation. The firing strength is given by

$$w_j = \prod_{k=1}^p \mu_{A_{kj}}(Z_{t-k}) \tag{66}$$

The third layer calculates normalised firing strength based on sum of all the other rules' firing strengths.

$$\overline{w_j} = \frac{w_j}{\sum_{j=1}^m w_j} \tag{67}$$

In the fourth layer, a parameter function is calculated by adaptive nodes based on the output of layer 3. These are called consequent parameters.

$$\overline{w}_j Z_t^{(j)} = \overline{w}_j(\theta_{j1}Z_{t-1} + \theta_{j2}Z_{t-2} + \cdots + \theta_{jp}Z_{t-p} + \theta_{j0}) \tag{68}$$

In the fifth layer, a single node exists in the fifth layer that aggregates all the incoming output signals from layer 4 and performs summation.

$$Z_t = \sum_{j=1}^m \overline{w}_j(\theta_{j1}Z_{t-1} + \theta_{j2}Z_{t-2} + \cdots + \theta_{jp}Z_{t-p} + \theta_{j0}) \tag{69}$$

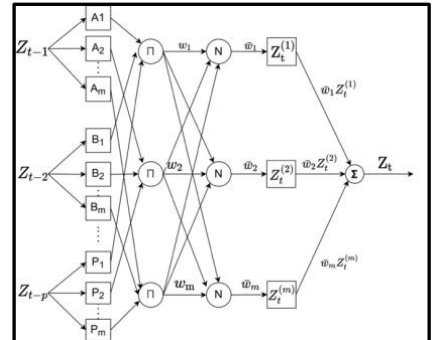

Fig. 6. Structure of a 5-layer ANFIS

### 3.4. PI&PID:

The algorithm used for a typical PI and PID can be shown as

$$u(t) = K_p e(t) + K_i \int e(t)dt \tag{70}$$

$$u(t) = K_p e(t) + K_i \int e(t)dt + K_p \frac{de}{dt} \tag{71}$$

where, $u(t)$ is the PID control variable, $K_p$ is the proportional gain, $e(t)$ is the error value, $K_i$ is the integral gain, $\frac{de}{dt}$ is the change in error with time. Changes in these constants impact the plant to various degrees, and hence PI and PID controllers undergo tuning where the constants are varied as per requirement. The addition of the derivative gain $K_p$ in the PID control loop ensures that the system is ready to handle rapid changes in setpoint values at any given point in time. Therefore, although more complex than PI, PID controllers offer more accuracy under unstable conditions and are generally more robust.

## 4. SIMULATION AND RESULTS

### 4.1. Implementation of simulated designs

Using extensive mathematical modelling and free body analysis, force equations were constructed and a Simulink subsystem was designed to accurately simulate the physical behaviour of an Inverted Pendulum on a Cart, shown in Fig. 7. The physical variables were taken as shown in table 1:

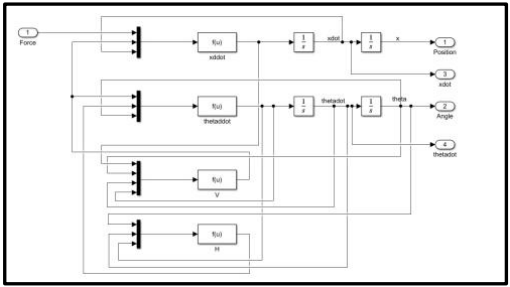

Fig. 7. Simulink subsystem designed for the Inverted Pendulum

Table 1. Physical variables and their values.

| Parameters | Values |
|---|---|
| mass of the cart $m_c$ | 0.5 kg |
| mass of the pendulum $m_p$ | 0.2 kg |
| coefficient of friction $b$ | 0.1 |
| moment of inertia of pendulum $J$ | 0.06 kg/m2 |
| gravitational acceleration $g$ | 9.8 m/s2 |
| length of the pendulum $l$ | 0.3 m |

The resulting transfer functions of the cart and the pendulum are calculated respectively as:

$$\frac{X(s)}{U(s)} = \frac{(4.182e-06)s^2 - (1.25e-04)}{(2.3e-06)s^4 + (4.182e-07)s^3 - (7.172e-05)s^2 - (1.025e-05)s} \quad (72)$$

$$\frac{\theta(s)}{U(s)} = \frac{(1.045e-05)s}{(2.3e-06)s^3 + (4.182e-07)s^2 - (7.172e-05)s - (1.025e-05)} \quad (73)$$

Likewise, the state transition matrix A and input matrix B take the form of:

$$A = \begin{pmatrix} 0 & 1 & 0 & 0 \\ 0 & -1.818 & 2.673 & 0 \\ 0 & 0 & 0 & 1 \\ 0 & -0.4545 & 31.18 & 0 \end{pmatrix} \quad (74)$$

$$B = \begin{pmatrix} 0 \\ 1.818 \\ 0 \\ 4.545 \end{pmatrix} \quad (75)$$

#### 4.1.1. LQR design:

The Q matrix of LQR is taken as,

$$Q = \begin{bmatrix} 1200 & 0 & 0 & 0 \\ 0 & 0 & 0 & 0 \\ 0 & 0 & 100 & 0 \\ 0 & 0 & 0 & 0 \end{bmatrix} \quad (76)$$

The (1,1) position of the Q matrix represents the cart position $x$ and the (3,3) position represents angle of the pendulum $\theta$. Therefore, we are prioritizing $\theta$ by penalizing control output aimed at $x$ about 12 times more. The aggression of the control action is set as $R = 1$.

#### 4.1.2. ANFIS design:

The parameters and factors used for training ANFIS in MATLAB are given in table 2. The resulting ANFIS structure is shown in Fig. 8.

Table 2. Parameters used for ANFIS design

| Parameters | Values |
|---|---|
| No. of inputs | 4 |
| No. of outputs | 1 |
| Membership functions | 2 |
| No. of datapoints | 500 (training), 91 (testing) |
| No. of epochs | 50 |
| Error tolerance (%) | 0 |
| Error obtained (%) | 0.0443 |
| No. of fuzzy rules | 16 |

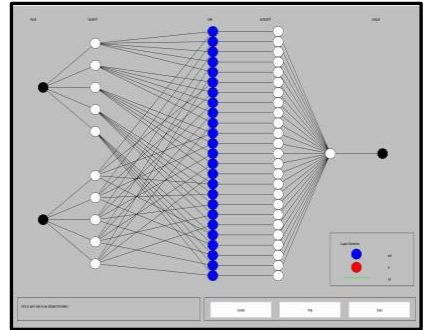

Fig. 8. Trained ANFIS structure

#### 4.1.3. PI & PID design:

The gains of the PI and PID were tuned for most optimal closed loop response curve, as given in table 3 and 4.

Table 3: Parameters Used For PI Design

| Parameter | Value |
|---|---|
| Proportional Gain Kp | 27.234 |
| Integral Gain Ki | 85.597 |

Table 4. Parameters Used For PID Design

| Parameter | Value |
|---|---|
| Proportional Gain Kp | 36.887 |
| Integral Gain Ki | 165.496 |
| Derivative Gain Kd | 1.505 |
| Filter coefficient N | 678.646 |

The two stages of TS-LA and PID control structures are implemented in Simulink as shown in Fig. 9, Fig. 10

and Fig. 11 respectively. A voltage to force conversion block is also implemented to simulate motor actuator effort.

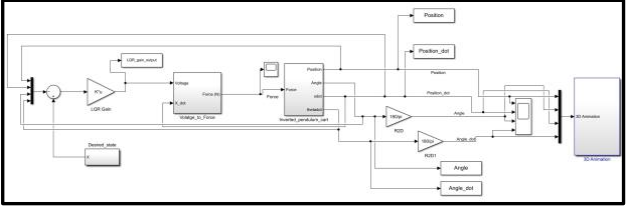

Fig. 9. Simulink implementation of the LQR training scheme

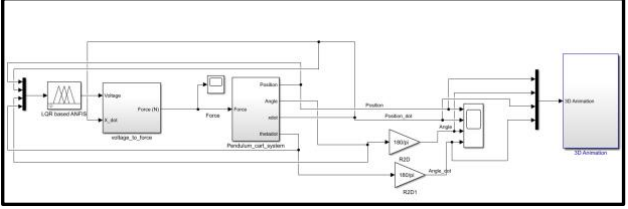

Fig. 10: Simulink implementation of the ANFIS control scheme.

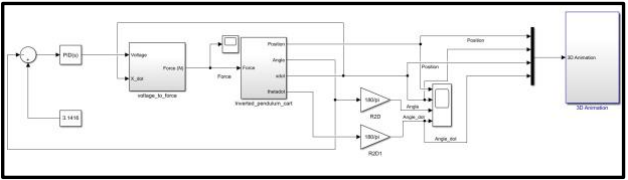

Fig. 11: Simulink implementation of the PID control scheme.

### 4.2. Testing using disturbance input

The inverted pendulum initially starts with all of its outputs at the unstable equilibrium position, implying the pendulum is balanced on its pivot upright. An impulse disturbance is given at around t=20s, which simulates a sharp knock given to the pendulum in the real world. The impulse disturbance input in Simulink is constructed using two step inputs cascading with each other. The settling time, which is the time taken for all oscillations to cease and the system to return to complete steady state, and angle of deviation of the pendulum upon impulse are considered for comparative analysis between the controllers.

### 4.3. Testing under white noise

Starting with the inverted pendulum in equilibrium, the system is subjected to continuous white noise disturbance which simulates crosswind effect and shear force in the real world as a source of uncertainty. The addition of such a disturbance represents cumulative macroscopical effect (output) of very high frequency fluctuations at the microscopic level (input).

### 4.4. Results and comparative analysis

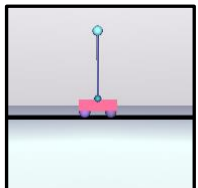

Fig. 12. Real time 3D simulation

In all the output graphs (Fig. 13, Fig. 14, Fig. 15 and Fig. 16, Fig. 17, Fig. 18), the first waveform represents cart position $x$, the second waveform represents cart velocity $\dot{x}$, the third waveform represents angle of the pendulum $\theta$, and the fourth waveform represents angular velocity of the pendulum $\dot{\theta}$. The entire system is visualised as shown in Fig. 12 using 3D CAD modelling and the control process is viewed in real time in MATLAB.

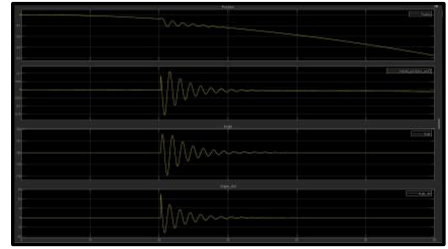

Fig. 13. Output waveforms from PI control

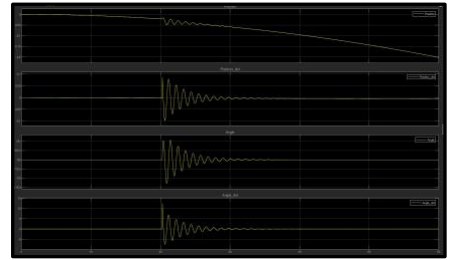

Fig. 14: Output waveforms from PID control

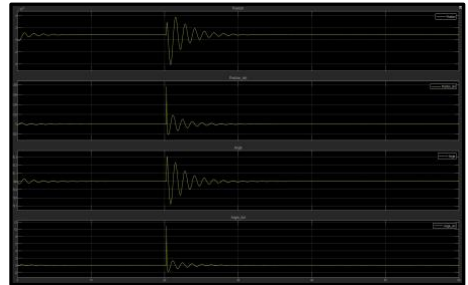

Fig. 15. Output waveforms from TS-LA control

During PID control the pendulum is deflected about 25% more compared to ANFIS control. Fig. 13, 14 and 15 show that with respect to the PID, the settling time is also improved by 22%, reducing cyclic oscillations during the process and therefore reducing wasted energy on controller effort. The system is much quicker to respond to changes as well, significantly by about 53.6% as shown by the rise time values. The experiment is repeated two more times with impulse magnitude constants 20 and 30 respectively and the mean transient values are recorded as shown below.

Table 4. Comparative analysis of various parameters

| Parameters | PI | PID | TS-LA |
|---|---|---|---|
| Settling time ($T_s$) | 17.25s | 12.5s | 9.75s |
| Deviation of $\theta$ | 2° | 1.2° | 0.3° |
| Rise time ($T_r$) | 107.19 ms | 68.73ms | 36.84ms |
| Steady state error ($\theta$) | 0 | 0 | 0 |
| Steady state error ($x$) | ∞ | ∞ | 0 |

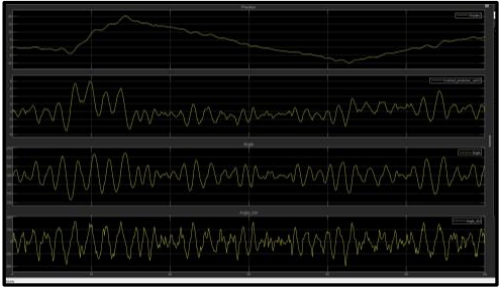

Fig. 16. Output waveforms from PI

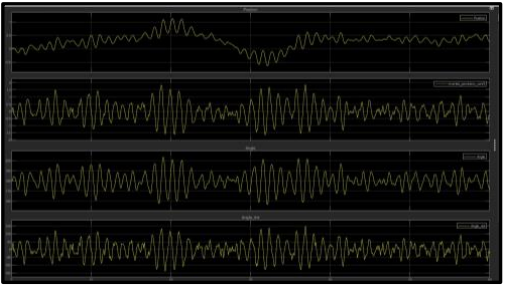

Fig. 17. Output waveforms from PID

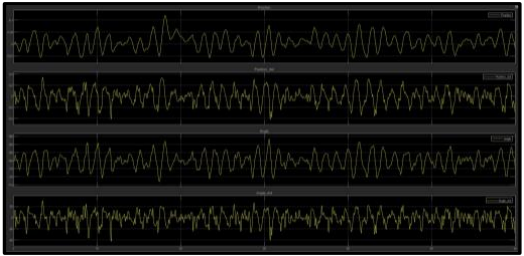

Fig. 18. Output waveforms from TS-LA

From Fig. 16, 17, and 18 we observe that when a constant source of white noise is given to the open loop system, the proposed controller achieves much better results. Significant improvements are seen in terms of mean/max deviation of the angle compared to the PI and PID controllers in terms of 78% and 45% respectively which shows that the TS-LA can mitigate instability within a closer margin. The maximum cart velocity $\dot{x}$ which is indicative of controller effort exerted on the actuator is improved by 95.25% and 78.9% when compared to PI and PID.

Table 5. Comparative analysis of various parameters

| Parameters | PI | PID | TS-LA |
|---|---|---|---|
| Max. deviation of $\theta$ | 50° | 20° | 11° |
| Max. deviation of $\dot{x}$ | 8 units | 1.8 units | 0.38 units |

Moreover, on obtaining the transient parameters from the output waveforms of the system, an important observation is made. As the PI and PID controllers are SISO, only one process variable can be controlled at a time. Therefore, only $\theta$ is under control while $x$ is not. However, due to the MISO nature of TS-LA, we are able to stabilize multiple variables at their respective equilibrium points as shown in Fig. 15 and 18. Hence the reason why the steady state error in terms of cart position during PI and PID control grows unbounded.

## 5. CONCLUSION

In this paper, the physical behaviour of an inverted pendulum on a cart system was accurately simulated with mathematical and state space modelling using Newtonian and Lagrangian mechanics. The system in open loop condition was proven unstable and controllable. Following this, a two-stage methodology was studied incorporating the advantages offered by LQR, Fuzzy Logic, and ANN to create a control structure (TS-LA) responsible for governing the motion of the system. Using impulse signals and white noise, the disturbance rejection processes was visualised in real time through 3D models and the outputs waveforms taken from TS-LA were benchmarked against standard conventional control, demonstrating the performance superiority and advantages of the proposed control technique. The control of an inverted pendulum is not the only application of TS-LA and other such advanced control techniques. They find their significance in industries like food and beverage, where the quality of the product is dependent on the control system being robust. The ability of being able to overcome nonlinear dynamics, time-delays and instabilities without much effort is highly sought after. It is important to note that laboratory and literature results may always vary from actual test case scenarios with real-time dynamics in play. More in-depth studies and implementations of said intelligent control techniques and their revised versions are therefore required, in conjunction with platforms like IIoT and plant monitoring systems.

## APPENDIX A

## CONFLICT OF INTEREST

The authors declare that there is no competing financial interest or personal relationship that could have appeared to influence the work reported in this paper.